\documentclass[%
reprint, amsmath,amssymb,
aps,
prl,
superscriptaddress,
]{revtex4-1}
\usepackage{rotating}
\usepackage{diagbox}
\usepackage{graphicx}
\usepackage{dcolumn}
\usepackage{bm}

\usepackage{ulem}

\usepackage[makeroom]{cancel}
\usepackage{dsfont}
\usepackage[english]{babel}
\usepackage{overpic}
\usepackage{amsmath}
\usepackage{bbold}
\usepackage{amsfonts}
\usepackage{braket}
\usepackage{amssymb}

\usepackage{xcolor}

\usepackage{makeidx}
\usepackage{bm}
\usepackage{dsfont}
\usepackage{graphicx}
\usepackage{kpfonts}
\usepackage[left=2cm,right=2cm,top=2cm,bottom=2cm]{geometry}
\usepackage{hyperref}
\hypersetup{
    colorlinks=true,
    linkcolor=blue,
    filecolor=magenta,      
    urlcolor=blue,
    citecolor=red,
}

\newcommand{\ba}{\begin{eqnarray}}
\newcommand{\ea}{\end{eqnarray}}

\begin{document}

\date{\today}
\begin{abstract}
The possibility of discriminating  the statistics of a thermal bath using indirect measurements performed on quantum probes is presented.
  The scheme relies on the fact that, 
 when weakly coupled with the environment of interest,
   the  transient evolution of the probe  toward its final thermal configuration, 
is strongly affected by the fermionic or bosonic nature of the bath excitations.
Using figures of merit taken from quantum metrology such as the Holevo-Helstrom probability of error and the Quantum Chernoff bound,
we  discuss how to achieve the greatest precision in this statistics tagging procedure, analyzing different models of probes and different initial preparations and by optimizing over the time of exposure of the probe. 
\end{abstract}
\title{Quantum bath statistics tagging} 
\author{Donato Farina}
\affiliation{NEST, Scuola Normale Superiore and Istituto Nanoscienze-CNR, I-56126 Pisa, Italy}
\affiliation{Istituto Italiano di Tecnologia, Graphene Labs, Via Morego 30, I-16163 Genova, Italy}
\author{Vasco Cavina}
\affiliation{NEST, Scuola Normale Superiore and Istituto Nanoscienze-CNR, I-56126 Pisa, Italy}
\affiliation{Complex Systems and Statistical Mechanics, Physics and Materials Science Research Unit, University of Luxembourg, L-1511 Luxembourg}
\author{Vittorio Giovannetti}
\affiliation{NEST, Scuola Normale Superiore and Istituto Nanoscienze-CNR, I-56126 Pisa, Italy}
\maketitle

In equilibrium statistical mechanics, the intrinsic indistinguishability between identical particles gives rise to the Bose-Einstein and Fermi-Dirac 
equilibrium distributions.
These statistics found their earliest evidences in matter physics, describing black body radiation \cite{Bose1924} and the behavior of electrons in solids \cite{sommerfeld1927} while their link with the intrinsic angular momentum of elementary particles stems as a crucial result of quantum field theory \cite{pauli-spin-statistics, schwabl2008advanced}.
A standard tool to discern the statistics of a quantum system is represented by 
two-body correlations, experimentally accessible through equilibrium response properties to  weak external fields \cite{mahan2013many}.
For example typical and exclusive signatures are the Pauli hole in case of fermions \cite{giuliani2005quantum} and bunching and anti-bunching phenomena in case of bosons \cite{paul1982photon}.
More in general, statistics tagging turns out to be worth in all modern physics. For instance, in astrophysics, methods  to recognize the statistical distributions of particles which are thermally radiated by black holes have been developed \cite{black-hole} or, going beyond \textit{conventional} fermions and bosons, in the context of the fractional quantum hall effect \cite{laughlin1983anomalous} interferometric measurements \cite{goldman2005fractional} confirmed the existence of quasi-particles obeying fractional exclusion statistics
\cite{wilczek1982quantum, haldane1991fractional, wu1994statistical}.
Finally,  from a technological point of view,
a detailed characterization of the environment surrounding a quantum system is nowadays crucial to implement quantum information protocols
and, more generally, for quantum nanotechnology   \cite{palma1996quantum, nielsen2002quantum}.  
Indeed, the interaction with the environment  leads to decoherence and dissipation on the system, strongly degrading purely quantum resources \cite{nielsen2002quantum} or even, in other cases, promoting collective quantum phenomena \cite{scarlatella2016dissipation}.

\begin{table}[t]
\begin{center}
\begin{tabular}{ |c|c|c| } 
 \hline
 \diagbox{Probe $A$}{Bath $B$
 }
   & fermionic & bosonic \\
   \hline 
 TLS & $\gamma$ & $n_{\rm th}\gamma$  \\ 
 \hline
 QHO & $\gamma/n_{\rm th}$ & $\gamma$ \\ 
 \hline
\end{tabular}
\end{center}
 \caption{Transition rates 
 governing the dynamics of the system-bath models for the four
 scenarios considered in the paper: in this expressions $\gamma$ is a constant that only depends upon the interaction strength of the model, while    $n_{\rm th}$ depends on $\beta$  as in Eq.~(\ref{eq:nth-definition}). Notice that 
 for homogeneous settings  (TLS-fermions or QHO-bosons) the values of the rates
 are independent from the 
 bath temperature. Furthermore since $n_{\rm th}\geq 1$   we observe that for the TLS probe the transition rate
 associated with the bosonic bath is always larger than the corresponding fermionic value, while exactly the opposite occurs for the QHO probe scenario.
 We  also recognize that in both the TLS and QHO configuration, the difference between the  transitions rates induced by 
the bosonic and fermionic statistics 
increases with the temperature. Such gap  nullify instead in the zero-temperature limit ($\beta\rightarrow \infty$) where 
$n_{\rm th}=1$: accordingly under this conditions the dynamics  of the model is expected not to
detect any difference in the bath statistics.
}
 \label{table:decay-rates}
 \end{table}

The characterization of measurement processes and statistical inference methods applied to quantum systems is the core of quantum metrology \cite{paris2009quantum, giovannetti2006quantum, giovannetti2011advances}.
The estimation and the discrimination of environmental properties can be achieved both via direct measurements or indirectly, by extracting information from auxiliary systems.
For instance, via putting a probe in contact with a thermal environment and performing a measure on such a probe, it is possible to extract information about the temperature \cite{correa2015individual,NATCOM,ANTO1,Cavina2018} and the spectral properties \cite{salari2019quantum, benedetti2018quantum} of the environment itself. 
Following this line of reasoning, we present a protocol aimed to discriminate between fermionic and bosonic thermal baths via indirect quantum state discrimination on an auxiliary quantum probe $A$.
More precisely  in our construction the tagging of the bath statistics is performed by 
monitoring the state of $A$ at a convenient finite time evolution $\bar{t}$ during the thermalization process
it experiences once put in weak-coupling thermal contact~\cite{breuer2002theory}  with the environment. 
The scheme ultimately relies on the fact that, while the final configuration of $A$ is not necessarily 
influenced by the statistical nature of the bath, 
the latter leaves residual imprintings on 
the transient of the thermalization process which can be picked up by proper measurements on the probe.
A full characterization of the ultimate discrimination efficiency we can achieve using this technique will 
be presented by studying a couple of paradigmatic examples where $A$ is assumed to be either
 a two level system (TLS) or a quantum harmonic oscillator (QHO).
It is worth stressing that the resulting four scenarios describe situations which are routinely encountered in experiments \cite{farina2019open} paving the way for a proof of principle implementations of our findings (at least): 
indeed 
a two level system coupled to a bosonic bath $(TLS-bosons)$ is paradigmatic in quantum optics  \cite{walls2007quantum} and quantum computation \cite{palma1996quantum}; 
a harmonic oscillator interacting with a bath of other harmonic oscillators $(QHO-bosons)$ can describe an open opto-mechanical resonator \cite{groblacher2015observation};
finally, spin-baths are more rare but also feasible \cite{pekola2016finite} if we deal with a vibrational degree of freedom interacting with two-level defects $(QHO-fermions)$ in quantum-electromechanical systems \cite{blencowe2004quantum, schlosshauer2008decoherence} 
or with the hyperfine interaction of an electron spin in a quantum dot with the surrounding nuclear spins $(TLS-fermions)$ \cite{prokof2000theory, urbaszek2013nuclear, fischer2018signatures, bortz2007exact}.
\paragraph{The model:--}
Let $B$ be a thermal bath characterized by a temperature $1/\beta$ that for simplicity we assume to be known. Our goal is to determine the statistical nature
 of the excitations of $B$ which is taken to be either bosonic or fermionic. For this task
 we are allowed to initialize the quantum probe $A$ in some fiduciary state $\rho(0)$, put it into thermal coupling with 
$B$ and then monitoring its final state after some interaction time $t$ as being elapsed.  
In our analysis we shall describe the associated dynamical evolution of $A$ by assigning  
a Gorini-Kossakowski-Sudarshan-Lindblad 
\cite{lindblad1976generators, gorini1976completely} master equation (ME) defined by the properties of $B$.
Independently from our choice of using a TLS probe or a QHO probe 
 $A$, the main feature that enables to distinguish between the actions of a fermionic and a bosonic bath is 
the time scale of the associated thermalization event. Indeed, 
as follows from the standard {Born-Markov-Secular microscopic derivation} of the ME (see~\cite{SUPMAT} for details), 
the  transition rate 
 associated with a given energy level spacing $\omega_0$ of $A$ 
can be expressed as shown in Table~\ref{table:decay-rates},
with $n_{\rm th}$ being the ratio 
between the associated Bose-Einstein ($N_b(\beta):=1/({e^{\beta \omega_0}-1})$) and the Fermi-Dirac 
 ($N_f(\beta):=1/({e^{\beta \omega_0}+1})$) occupations numbers,
i.e. the quantity 
\begin{eqnarray} \label{eq:nth-definition}
n_{\rm th}:={N_b(\beta)}/{N_f(\beta)}=\coth[\beta 
\omega_0/2]\;, \end{eqnarray} 
(hereafter $\hbar=1$).
 Based on this observation we can hence translate  the two possible choices for $B$ 
 into two possible 
 hypotheses $\rho_b(t)$ and $\rho_f(t)$ for the density matrix $\rho(t)$ at a certain time $t$,
corresponding, respectively, to the evolved state of $A$ via the bosonic and the fermionic thermal channels.

\begin{figure}
\vspace{1cm}
\begin{overpic}[width=0.49\linewidth]{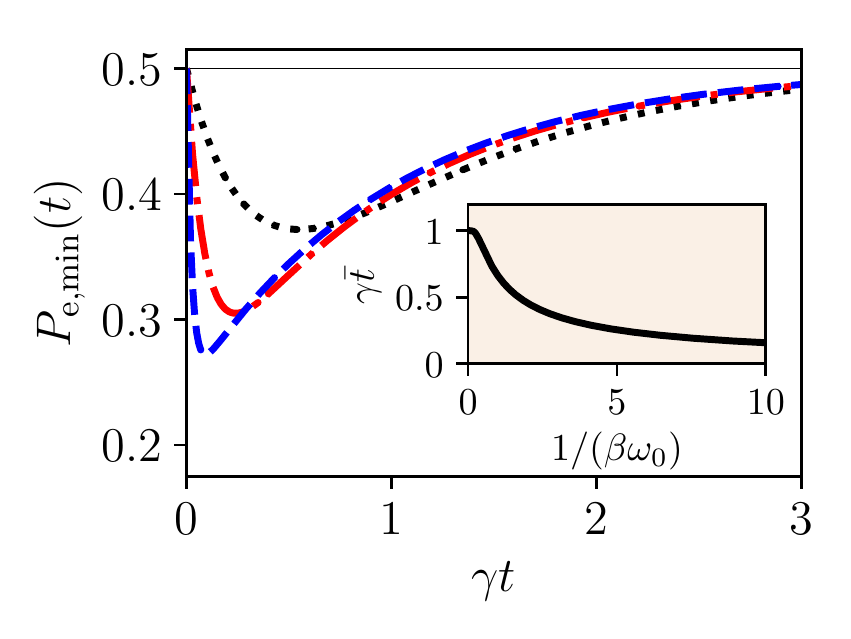}\put(10,75){(a)} \end{overpic}
\begin{overpic}[width=0.49\linewidth]{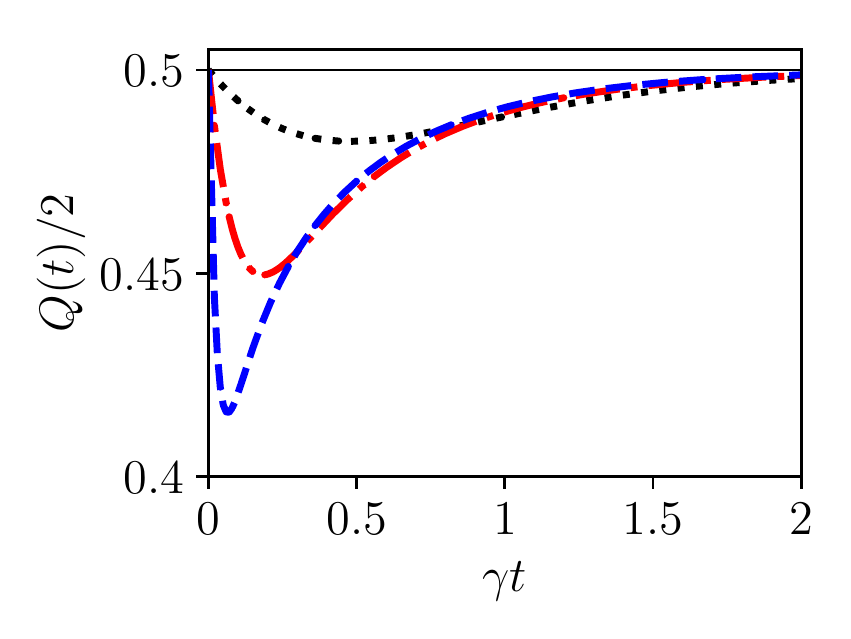}\put(10,75){(b)} \end{overpic}
\caption{
\label{fig:gnd0}
(Color online)
Plots of the (a) Helstrom error probability (\ref{Helstrom}), 
 and (b) the (rescaled) Chernoff  quantity~(\ref{eq:chernoff-bound})   for the TLS probe case, initialized in the excited state, as a function of the measurement time $t$. The three curves 
represent three different bath temperatures: $1/(\beta \omega_0) = 1.5 $ (black dotted line),  $1/(\beta \omega_0) = 5.5 $ (red dot-dashed line) and $1/(\beta \omega_0) = 20.5 $ (blue dashed line). 
The inset in (a) shows $\gamma \bar{t}$ as function of $1/(\beta \omega_0)$ for the Helstrom error probability.
} 
\end{figure}

In general, the 
discrimination between two quantum states involves a measurement process.
If we choose wisely the measurement and the successive inference procedure, we will be able to discriminate between the two hypotheses with the highest precision.
A natural quantifier of the effectiveness of such a method
is given by $1 - P_{e}$, where $P_e$ is the error probability, that is the probability to guess incorrectly the state after reading the measurement outcomes.
In the two state discrimination problem 
$P_{e}$ has been minimized over all the set of possible measurements protocols by 
Helstrom and Holevo \cite{helstrom1976,holevo1973}.
This optimal value quantifies how much two quantum states, for instance 
our $\rho_f(t)$ and $\rho_b(t)$,
are distinguishable:
\begin{equation} \label{Helstrom} P_{\rm{e,min}}(t) :=
\frac{1}{2}
 \left( 1 - \frac{1}{2} \| \rho_{b}(t)-\rho_{f}(t) \|_1\right)~,  \end{equation}
where $\| \cdot \|_1$ denotes the trace norm. 
More generally if we have $N\geq 1$ identical probes at disposal, the discrimination process involves $\rho_{b}(t)^{\otimes N}$ and $\rho_{f}(t)^{\otimes N}$ while the minimum probability of error 
satisfies

\begin{equation} 
\label{eq:error-chernoff} P^{(N)}_{\rm e,min}(t) \leq 
 {Q(t)^N}/{2}, \end{equation}
where 
$Q(t)$ is minimum of the Chernoff function $Q_r(t)$, i.e. 
\begin{eqnarray}
\label{eq:chernoff-bound}
Q(t)=\min_{r \in [0, 1]} Q_r(t)~,
\qquad Q_r(t):={\rm tr}\left[ \rho_b^r (t) \rho_f^{1-r} (t) \right]~.
\end{eqnarray}
The result (\ref{eq:error-chernoff}) is known as Quantum Chernoff Bound 
\cite{ogawa2004error} and is asymptotically tight
for $N \rightarrow \infty$ ~\citep{audenaert2007discriminating}.
Both 
the quantities defined in Eq. (\ref{eq:chernoff-bound}) and
Eq. (\ref{Helstrom}) provide operationally well defined figures of merit for the precision in the discrimination between 
$\rho_b(t)$ and $\rho_f(t)$.
In what follows we shall analyze their dependence  from the initial state  of the probe and perform a further minimization with respect to $t$
to determine the best time instant $\bar{t}$ for the quantum state discrimination.
\paragraph*{\noindent Statistical tagging via TLS probe: --}\label{sec-qubit}
Here we present a complete analysis of the problem for the case where $A$ is a TSL
with local Hamiltonian $H = \omega_0 \sigma_+ \sigma_- $, $\sigma_\pm$ being the associated ladder operators. The corresponding ME  
induced by a bosonic/fermionic environment is~\cite{SUPMAT}  
\begin{eqnarray}
\label{eq:mastertls}
&\dot{ \rho}_q(t)=
-i [H, \rho_q(t)]+\gamma N_q(\beta) \left(\sigma_+ \rho_q(t) \sigma_- - \frac{1}{2} \{\sigma_- \sigma_+ , \rho_q(t) \}\right)&\nonumber\\
&+\gamma [ 1 + s_q N_q (\beta) ] \left(\sigma_- \rho_q(t) \sigma_+ - \frac{1}{2} \{\sigma_+ \sigma_- , \rho_q(t) \}\right)
~,&
\end{eqnarray}
where $q\in \{ b,f\}$ and $s_b=1$ and $s_f=-1$.
In the Bloch coordinates representation
$\rho_{q}(t) = \frac{1}{2} ( \mathds{1} + \vec{ \langle \sigma(t) \rangle}_{q} \cdot 
\vec{\sigma})$ an integration of Eq. (\ref{eq:mastertls}) results in  $\langle \sigma_z(t)\rangle_{q} = \langle \sigma_z(0)\rangle e^{-\gamma n_{\rm th}^{(q)} t} + (1- 2 N_f(\beta))( e^{-\gamma  n_{\rm th}^{(q)} t} -1)$, and $\langle\sigma_x(t)\rangle_q = \langle \sigma_x(0) \rangle 
e^{-{\gamma n^{(q)}_{th} t }/{2}}$,
where
\begin{eqnarray} n_{\rm th}^{(q)}:= n_{\rm th} + (1-s_{q})(1-n_{\rm th})/2,
 \end{eqnarray} 
  is the TLS rate renormalization factor given in Table~\ref{table:decay-rates}, and
where $\langle \sigma_{x,z}\rangle(0)$ are the initial conditions ($\langle \sigma_y(0) \rangle$ being set equal to $0$ exploiting the $x-y$ symmetry of the problem).
At $t= + \infty$, the probe  thermalizes at the equilibrium values $\langle\sigma_z \rangle_{eq}= 2 N_f(\beta) -1$, $\langle\sigma_x \rangle_{eq} = 0$ irrespectively from the bath statistics (i.e. $\rho_{b}(\infty) = \rho_{f}(\infty)$),
implying 
that  the discrimination between the bosonic and fermionic environments becomes impossible. For this reason, the measurement time $\bar{t}$ will be finite and can be found by maximizing the trace distance between $\rho_b(t)$ and $\rho_f(t)$ according to Eq. (\ref{Helstrom}).
The best discriminating strength is obtained by
initializing $A$ in the excited state  of its local Hamiltonian (i.e. $\langle \sigma_z(0)\rangle = 1$, $\langle \sigma_x(0)\rangle = 0$)~\cite{SUPMAT}.
Intuitively, such input configuration is the farthest from the equilibrium configuration, and this choice allows the  ``faster" bosonic thermalizing probe to outdistance on a longer track its fermionic counterpart, increasing their distinguishability.
In particular, plugging $\langle \sigma_z(0)\rangle = 1$, and $\langle \sigma_x(0)\rangle = 0$  we get 
 $|| \rho_b(t) - \rho_f(t) ||_1 = (1-\langle \sigma_z \rangle_{eq}) 
  (e^{-\gamma n^{(f)}_{th} t} - e^{-\gamma  n^{(b)}_{th} t})$
whose associated value of $P_{\rm{e,min}}(t)$ is reported in Fig. \ref{fig:gnd0}(a)  for different choices of the bath temperatures. As anticipated in the limit of large time $t$ the error asymptotically approaches $1/2$ indicating
the failure of the tagging procedure. Minimum values for  $P_{\rm{e,min}}(t)$ are instead obtained for an optimal choice
of $t$ given by 
\begin{equation} \label{eq:bartqub} \bar{t} =  { \log (n_{\rm th}) }/(2{ \gamma N_b(\beta)}) =
{ \log (n_{\rm th}) }/({ \gamma (n_{\rm th} -1)}) \;,
 \end{equation}
whose functional dependence upon $\beta$ is reported in the inset of the figure. 
As anticipated in the caption of Table~\ref{table:decay-rates} the model exhibit no discrimination strength at zero temperature
where $P_{\rm{e,min}}(t)=1/2$, while 
 better discriminating strength is achieved at high temperatures since in this case $n_{\rm th}$ diverges, and so does the gap between  the bosonic and fermionic thermalization rates. 
Analogous  conclusions can be obtained also in the case where we have $N$ copies of the evolved state of the probe. Here  
exploiting the results of Ref.~\cite{calsamiglia2008quantum} the functional $Q_r(t)$ can be computed as
\begin{equation} \label{eq:chernq} \begin{gathered} Q_r(t) = [\lambda_b^r \lambda_f^{1-r} + (1-\lambda_b)^{r} (1- \lambda_f)^{1-r}] \cos(\tfrac{\theta}{2})^2 \\
+ [\lambda_b^r (1- \lambda_f)^{1-r} + (1- \lambda_b)^r \lambda_f^{1-r}] \sin(\tfrac{\theta}{2})^2,  \end{gathered} \end{equation}
where $\lambda_{q}$ is the greatest eigenvalue of $\rho_{q}(t)$ and $\theta$ is the angle between the Bloch vectors associated to $\rho_f(t)$ and $\rho_b(t)$. By numerical optimization with respect to $r$ the resulting value of $Q(t)$  are reported 
 in Fig. \ref{fig:gnd0}(b), and 
qualitatively provides the same insight we obtained  from the Helstrom probability analysis. 
Notice that in both cases, 
we have  crossing between curves associated, meaning that if we wait too much 
(and lose the opportunity of measuring in $\bar{t}$) 
the discrimination becomes easier at low temperatures (this property will not occur when probing with a QHO, as we will see in the next paragraph).

\begin{figure}
\vspace{1cm}
\begin{overpic}[width=0.49\linewidth]{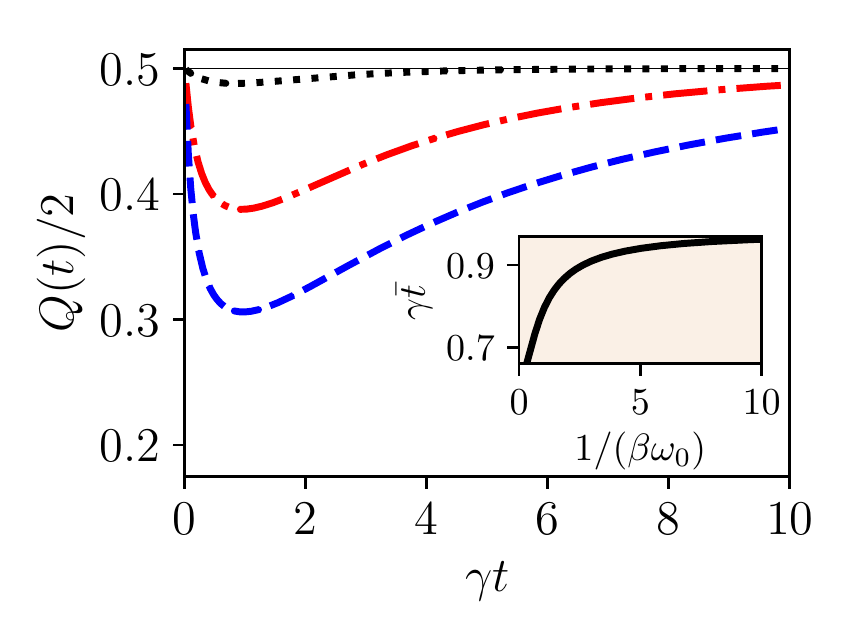}\put(10,75){(a)} \end{overpic}
\begin{overpic}[width=0.49\linewidth]{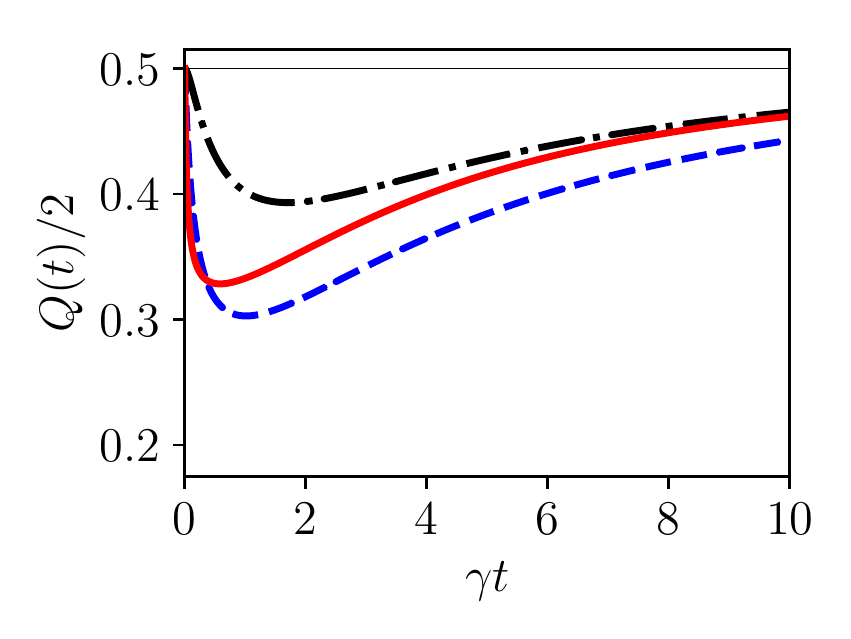}\put(10,75){(b)} \end{overpic}
\caption{
\label{fig:ho}
(Color online) (a): Plot of the  (rescaled) Chernoff quantity (\ref{eq:chernoff-bound})
of the QHO model associated with the ground state input 
for $1/(\beta \omega_0) = 1.5 $ (black dotted line), $1/(\beta \omega_0) = 5.5 $ (red dot-dashed line) and $1/(\beta \omega_0) = 10.5 $ (blue dashed line).
The inset shows the associated $\gamma \bar{t}$ as function of $1/(\beta \omega_0)~.$
(b): 
Plot of $Q(t)/2$ for different choices of the initial Gaussian state for fixed initial mean excitation number $\langle a^\dag a (0)\rangle =1$ 
(for $1/(\beta \omega_0)=10.5$): 
coherent state (blue dashed line),   
thermal state (black dot-dashed line), 
squeezed ground state (red full line).
} 
\end{figure}

\paragraph{Statistical tagging via QHO probe:--}
\label{sec-qho}
Assume next the probe $A$ to be a QHO 
of  Hamiltonian 
$H=\omega_0 a^\dag a$, and 
evolving via the  ME 
\begin{eqnarray}
\label{eq:lindblad-eq}
&\dot{ \rho}(t)=
-i [H, \rho(t)]+
\gamma [ 1 + s_q N_q (\beta)] \left(a \rho(t) a^\dag - \frac{1}{2} \{a^\dag a , \rho(t) \}\right)&
\nonumber\\
&+\gamma N_q(\beta) \left(a^\dag \rho(t) a - \frac{1}{2} \{a a^\dag , \rho(t) \}\right)~,&
\end{eqnarray}
where $s_q$ and $N_q (\beta)$ are defined as  in 
Eq.~(\ref{eq:mastertls}).
An explicit integration of Eq.~(\ref{eq:lindblad-eq}) can be easily obtained
in the case of Gaussian input states~\cite{weedbrook2012gaussian,serafini2017quantum,ferraro2005gaussian}  
 having vehemently pursued experimental realizations (see e.g. \cite{laurat2005entanglement, adesso2014continuous}),  which can be expressed as displaced, squeezed thermal states of the form 
\begin{eqnarray} 
\rho(0)= D^\dag ({\pmb{\xi}}_0) S^\dag(\chi_0) \frac{e^{-\beta_0 H}}{Z(\beta_0)}  S(\chi_0)  D({\pmb{\xi}}_0)\;,  \label{INPUTGAUS} 
\end{eqnarray} 
$Z(\beta_0) := \mbox{Tr} [e^{-\beta_0 H}]$ being a normalization factor. 
The dynamics of these inputs is completely determined by the first and second momenta of the
system annihilation and creation operators which, by direct integration, yield the following expressions 
 $\langle a(t) \rangle=\langle a(0)\rangle  e^{-\frac{\gamma}{2}/n_{\rm th}^{\rm (q)} t} e^{- i  \omega_0 t}$, $\langle a^2(t) \rangle =\langle a^2(0) \rangle  e^{-\gamma/n_{\rm th}^{\rm (q)} t} e^{-2 i  \omega_0 t}$, and 
$\langle a^\dagger a (t)\rangle =\langle a^\dagger a (0) \rangle e^{-\gamma/n_{\rm th}^{\rm (q)} t}+ 
N_b(\beta) 
(1-e^{-\gamma/n_{\rm th}^{\rm (q)} t})$,
which exhibits a transition rate renormalization factor $1/n_{\rm th}^{\rm (q)}$ that is the inverse of the one observed for TLS model as anticipated in Table~\ref{table:decay-rates}.
We immediately notice that once more at zero temperature $(n_{\rm th}=1)$ the probe dynamics is insensitive to the bath statistics (as it was also clear from Eq.~(\ref{eq:lindblad-eq})). The same occurs for generic $\beta$ in 
the asymptotic limit $t\rightarrow\infty$ where, independently of the initial state and of the statistics of the bath 
the system obtains an average number of photons $\langle a^\dag a (\infty) \rangle = N_b(\beta)$ and the coherences disappear: $\langle a (\infty) \rangle= \langle a^2 (\infty) \rangle = 0~.$
As a measure of distinguishability of the associated $\rho_b(t)$ and $\rho_f(t)$ counterparts of
the input (\ref{INPUTGAUS}) 
we utilize the quantum Chernoff quantity~(\ref{eq:chernoff-bound}) for which 
a convenient formula for Gaussian states is known \cite{calsamiglia2008quantum, weedbrook2012gaussian}. A detailed account of this calculation is presented in~\cite{SUPMAT}: the obtained results are summarized in Fig.~\ref{fig:ho} for different choices of the input parameters.
In particular in panel (a) we plot the value of $Q(t)$ for the case in which $\rho(0)$
is the ground state of the QHO (i.e. ${\pmb{\xi}}_0=0$, $\chi_0=0$, and $\beta_0\rightarrow \infty$):
as in the TLS case we notice that the discrimination efficiency gets depressed in the asymptotic limit of sufficiently large evolution times $t$,
reaching a maximum value for intermediate values of the parameter. The performances
gets also affected by the value of the bath temperature, with higher sensitivity being attained for large values of $1/\beta$ (i.e. large values of $N_b(\beta)$). 
In Fig.~\ref{fig:ho}(b) instead we give a comparison of the performances obtained for different
choices of possible input states (coherent state, thermal state, squeezed ground state) characterized by an identical value of the  
 initial average number of photons $\langle a^\dag a (0)\rangle =1$.
As the plot shows, all cases exhibit the same functional dependence  observed for the ground state input. Nonetheless 
introducing the initial energy via displacement leads to the lowest error probability, while squeezing is effective for it to be attained in short time.
We also remark that in the absence of the input energy  limitation, $Q(t)$ can be brought to reach arbitrarily small values 
because of the possibility of injecting arbitrarily large initial energy into the system
(clearly an analogous effect cannot be found when probing the bath with a TLS due to the
limited Hilbert space of the latter).  
As a final observation we notice that closed analytical expressions that capture the above behaviours can be obtained  in the special case where the  initial state $\rho(0)$ is not squeezed and has a temperature that is identical to the bath temperature ($\beta_0=\beta$). It turns out that with this choices the resulting expression for $Q_r(t)$ is particularly compact
$Q_r(t)=\exp\left\{ 
-
\frac{|{\pmb{\delta}}(t)|^2}{2} 
\left[
1+2N_b(\beta)-N_b(\beta)
f_r
\right]
\right\}$,
with {$f_r:=\left(1+\frac{1}{N_b(\beta)}\right)^r+\left(1+\frac{1}{N_b(\beta)}\right)^{1-r}~$}
and $\pmb{\delta}(t) :={\pmb{\xi}}_0
\left(e^{-\frac{\gamma}{2} t}-e^{-{\frac{\gamma}{2} t}/{ n_{\rm th}}}\right)$. In this case
 the minimum of $Q_r(t)$ can be easily shown to be   attained for 
$
r=1/2
$.
As a result we get 
\begin{equation}
\label{eq:chernoff-bound-displ-same-beta0beta}
Q(t)=
\exp\left\{ 
-
\frac{1}{2} 
\left[
\sqrt{N_b(\beta)+1}-\sqrt{N_b(\beta)}
\right]^2
|\pmb{\delta}(t)|^2
\right\}~,
\end{equation}
which can now be optimized with respect to $t$ leading to the analytical expression
that mimics the one observed in the TLS analysis,
\begin{equation}
\label{eq:tbar-ho-analytic}
\bar{t}={ \ln(n_{\rm th})}/({\gamma N_f(\beta)})= 2 n_{\rm th}  \log (n_{\rm th})/({ \gamma (n_{\rm th} -1)})\;.
\end{equation}
Feeding this into Eq.~(\ref{eq:chernoff-bound-displ-same-beta0beta}) the
 resulting expression can now be optimized with respect to the bath temperature $\beta$, 
giving $N_b(\beta_{\rm best})\approx 1.96$ corresponding to values 
$\bar{t}_{\beta_{\rm best}}\approx 4 /\gamma$ 
and $Q(\bar{t}_{\beta_{\rm best}})=\exp(-\kappa\, |{\pmb{\xi}}_0|^2)$
with $\kappa\approx 0.0145$.
\paragraph{Conclusions:--}
We studied how to 
 tag the quantum statistics of a thermal bath in an indirect way, using an auxiliary probe and a quantum measurement scheme.
Upon optimizing over the initial state of the probe, such discrimination turns out to be feasible during the time transient, i.e. before thermalization.
The efficiency of the discrimination relies on 
the fact that in heterogeneous settings - TLS/bosonic bath, QHO/fermionic bath - the temperature renormalizes the thermalization rates.
This approach can lead to significant advances in the problem of the statistics tagging, which is central in several fields \cite{black-hole, goldman2005fractional, wilczek1982quantum, haldane1991fractional, wu1994statistical}.
Generalization of the present analysis include the possibility of using more sophisticated
techniques (such as Choi-Jamiolkowski or diamond norm discrimination procedures~\cite{WILDE,WATROUS}) aimed to directly tag the generators associated with different
bath statistic without focusing on special input states of the probe. 
 As a further developement we also notice that, with some minor variations, 
  the method proposed  can be easily adapted to the discrimination of non conventional statistics interpolating between fermions and bosons.
  
We finally conclude by stressing that the proposed scheme 
can clearly be considered as a subroutine to be used in conjunction with 
other already existing  probe-mediated quantum metrology schemes
to provide a complete reconstruction
of the
 bath properties that, beside statistical characterization of its excitations, will include also other relevant quantities like the temperature or bare thermalization rate.

\begin{widetext}
\newpage

\section{SUPPLEMENTAL MATERIAL}
\end{widetext}
\appendix

\section{Balance law and temperature dependent rates} \label{appendix-detailed-balance}
In what follows we shall adopt a compact notation that allows us to treat uniformly the four
possible scenarios, {\it TLS-bosons, TLS-fermions, QHO-bosons}, and {\it QHO-fermions}.
For this purpose we introduce a system annihilation operator $\zeta_p$
where the subscript $p\in \{ \mbox{TLS, QHO}\}$ refer to the two possible
species of probes, assuming that 
 $\zeta_{p=\mbox{QHO}}=a$ and $\zeta_{p=\mbox{TLS}}=\sigma_-~.$
With this choice the can now describe 
the coupling between $A$ and its environment $B$ by assigning the
microscopic Hamiltonian $H_{AB}=H+H_B+H_I$  characterized by the following components 
\begin{eqnarray}
H&=&\omega_0 \zeta_p^\dag \zeta_p~,\\
H_B&=&\sum_k \omega_k c_q^\dag(k)c_q (k)~,\\
H_I&=&\sum_k \gamma_k [c_q^\dag(k)+c_q (k)](\zeta_p+\zeta_p^\dag)~,
\end{eqnarray}
%
 where the environmental modes $c_q^\dag(k)$ and $c_q(k)$ can be either of bosonic $(q=b)$ or of fermionic nature $(q=f)$. Following the Born-Markov-Secular microscopic derivation for a thermal environment \cite{breuer2002theoryA}, 
the Lindblad equation for the four cases of interest can be written in a unified form as \cite{farina2019openA}
\begin{eqnarray}
\label{eq:lindblad-eq-general}
&\dot{ \rho}(t)=
-i [H, \rho]+\gamma N_q(\beta) \left(\zeta_p^\dag \rho(t) \zeta_p - \frac{1}{2} \{\zeta_p \zeta_p^\dag , \rho(t) \}\right)&\nonumber\\
&+\gamma [ 1 + s_q N_q (\beta) ] \left(\zeta_p \rho(t) \zeta_p^\dag - \frac{1}{2} \{\zeta_p^\dag \zeta_p , \rho(t) \}\right)
~,&
\end{eqnarray}
with $\gamma$ being the {bare} dissipation rate and with $N_q(\beta)$ being the bath mean excitation number corresponding to the frequency $\omega_0$ -- the input state of $B$ being assumed to be thermal with inverse temperature $\beta$. 
Equation~(\ref{eq:lindblad-eq-general})
implies the following balance equation for the mean excitation number ${\langle \zeta_p^\dagger \zeta_p (t) \rangle}$:
\begin{equation}
\label{eq:detailed-balance-2}
\frac{d}{dt}{\langle \zeta_p^\dagger \zeta_p (t) \rangle}=
-\gamma\left( {N_q}(\beta)/{N_p}(\beta)\right) {\langle \zeta_p^\dagger \zeta_p (t) \rangle} + \gamma N_q(\beta)
~,
\end{equation}
where we can recognize the characteristic rate $\gamma_{p-q}=\gamma {N_q(\beta)}/{N_p(\beta)}$ from which the result of Table I of the main text  follows automatically. 

To comment Eq.~(\ref{eq:detailed-balance-2}), let's consider a thermal charging, i.e. a system initially in its ground state gets excited by a finite temperature thermal bath, finally reaching the bath temperature $1/\beta$. A TLS interacting with a bosonic environment, realizes a situation in which the great amount of excitation contained in each QHO cannot be hosted by the TLS. This unbalance results in an increase of the charging rate. The opposite is expected to occur when a QHO interacts with a fermionic bath: increasing temperature is expected to decrease the charging rate. Finally, such effect must disappear at low temperature where the difference between the energy spectra is irrelevant, because $N_b(\beta) \sim N_f(\beta) \sim e^{-\beta \omega_0}$ for $\beta\rightarrow \infty$.

Other considerations about speed effects arising from coupling a system with a bounded spectrum and a system with an unbounded spectrum can be found in \cite{schlosshauer2008decoherenceA, farina2019chargerA, andolina2018chargerA}.

\section{Details on Gaussian states}\label{appendix-gaussians}
The most general single-mode Gaussian state can be expressed as a squeezed-displaced-thermal state of the form 
\begin{equation}
\label{eq:gaussian-state}
\rho_G({\beta}, {\pmb{\xi}}, \chi):= D^\dag ({\pmb{\xi}}) S^\dag(\chi)  \frac{e^{-{\beta} H}}{{\rm tr}\left[ e^{-{\beta} H}\right]}S(\chi) D({\pmb{\xi}})~. 
\end{equation}
In the above expression $\beta\geq 0$ defined the inverse temperature of the
 state, while  the complex parameter 
 $\chi$  and the 2-D real vector $\pmb{\xi}=(\xi_1,\xi_2)^T$  define 
the squeezing and the displacement operator respectively, i.e. 
\begin{eqnarray}
\label{eq:squeezing-op}
S(\chi)&=&\exp\left[\frac{1}{2} \left(\chi^* a^2 - \chi {a^\dag}^2\right)\right]\;, \\
%
\label{eq:displacement-op}
D({\pmb{\xi}})&=&\exp[-i({\xi}_2 x- {\xi}_1 y)]~,
\end{eqnarray} 
with the operators $x=(a+a^\dag)/\sqrt{2}$ and $y=(a-a^\dag)/(\sqrt{2}i)$ being the canonical 
quadratures of the model. 

\subsubsection{Displacement and squeezing} 
The displacement operator $D({\pmb{\xi}})$ of Eq.~(\ref{eq:displacement-op}) sets the first moments of the state (\ref{eq:gaussian-state}). Its action on the canonical variables is the following
\begin{equation}
D({\pmb{\xi}})~ \pmb{r}~ D^\dag({\pmb{\xi}})=\pmb{r}+{\pmb{\xi}}~,
\end{equation}
with $\pmb{r}:= \begin{pmatrix}
x\\
 y
\end{pmatrix}$.

The squeezing operator defined in Eq.~(\ref{eq:squeezing-op})
transforms the ladder operators $a$ and $a^\dag$ as follows  \cite{olivares2012quantumA, ferraro2005gaussianA}:
%
%
\begin{eqnarray}
&S(\chi)~ \pmb{a}~ S^\dag(\chi)=\mathbb{S}_A (\chi)~ \pmb{a}~,&
\\
&
\mathbb{S}_A(\chi):=
\begin{pmatrix}
\cosh(|\chi|) & e^{i 2 \phi} \sinh(|\chi|)\\
e^{-i 2 \phi} \sinh(|\chi|) & \cosh(|\chi|)
\end{pmatrix}~,
&
\end{eqnarray}
where $\pmb{a}:= \begin{pmatrix}
a\\
 a^\dagger
\end{pmatrix}$ and with $2\phi$ being the phase of $\chi$, i.e $\chi=|\chi| e^{i 2\phi}$. 
Alternatively this can also be expressed as 
\begin{eqnarray}
S(\chi)~ \pmb{r}~ S^\dag(\chi)=\mathbb{S}({\chi})~ \pmb{r}~, 
\end{eqnarray}
where now 
\begin{eqnarray}
&\mathbb{S}({\chi})=&\\
\nonumber
&
\begin{pmatrix}
\cosh(|\chi|) + \sinh(|\chi|) \cos(2 \phi) & \sinh(|\chi|) \sin(2 \phi)\\
 \sinh(|\chi|) \sin(2 \phi) & \cosh(|\chi|) - \sinh(|\chi|) \cos(2 \phi)
\end{pmatrix}
~,&
\end{eqnarray}
 the   matrices $\mathbb{S}({\chi})$ and 
$\mathbb{S}_A(\chi)$ being related via the transformation 
\begin{equation}
\mathbb{S}({\chi})=U \mathbb{S}_A (\chi) U^\dag~,
\end{equation}
with $U$ being the unitary matrix 
\begin{equation}
U=1/\sqrt{2} \label{UMAT} 
\begin{pmatrix}
1 & 1\\
-i & i
\end{pmatrix}
~. 
\end{equation}

\subsubsection{First and second moments of the Gaussian state} 

Define 
the vector 
\begin{equation}
\pmb{A}= \langle \pmb{a} \rangle = 
\begin{pmatrix}
\langle a\rangle\\
\langle a^\dagger\rangle
\end{pmatrix} \;,
\end{equation}
and the matrix 
\begin{eqnarray}
\sigma_{A}=
\begin{pmatrix}
2 \langle a^2\rangle - 2 \langle a\rangle^2&
 2 \langle a^\dagger a\rangle + 1 - 2|\langle a\rangle|^2 \\
2 \langle a^\dagger a \rangle + 1 - 2|\langle a\rangle|^2&
  \left[2 \langle a^2\rangle - 2 \langle a\rangle ^2\right]^*
\end{pmatrix}
%
~,
\nonumber
\end{eqnarray}
where $\langle ...\rangle$ represent the expectation value computed on the Gaussian state
of Eq.~(\ref{eq:gaussian-state}). 
From these expression one can then easily retrive the canonical first moments 
\begin{equation}
\pmb{R}= \langle \pmb{r} \rangle =\begin{pmatrix}
\langle x\rangle\\
\langle y \rangle
\end{pmatrix} \;,
\end{equation}
and the (real-symmetric) covariance matrix 
\begin{equation}
\label{eq:cov-mat-canonical}
\sigma_{ij}=\langle \{r_i- \langle r_i \rangle , r_j- \langle r_j \rangle   \} \rangle~,
\end{equation}
Indeed one has 
\begin{eqnarray}
\pmb{R}(t)=U \pmb{A}(t)~, \qquad \qquad 
\sigma(t) =U \sigma_A(t) U^T~, 
\end{eqnarray}
with $U$ as in Eq.~(\ref{UMAT}). 
From the above analysis it follows that 
 the moments of a Gaussian state (\ref{eq:gaussian-state}) hold
\begin{eqnarray}
\pmb{R}={\pmb{\xi}}~,\qquad  \qquad 
\label{eq:cov-mat-gaussian}
\sigma=\nu_{{\beta}} \mathbb{S}(\chi)  \mathbb{S}^T(\chi) ~,
\end{eqnarray} 
with
\begin{eqnarray}
\nu_{{\beta}}&=&2 N_b({{\beta}})+1=\coth\left({{{\beta}} \omega_0}/{2}\right)~.
\end{eqnarray} 
Equation~(\ref{eq:cov-mat-gaussian}) is better understood once it is written as 
\begin{eqnarray}
\sigma=\mathbb{S}(\chi)  \sigma_{{\beta}} \mathbb{S}^T(\chi) ~,
\qquad \qquad 
\sigma_{{\beta}}=\nu_{{{\beta}}}  \mathds{1}_2~,
\end{eqnarray}
where $\sigma_{{\beta}}$ is the covariance matrix of the thermal state ${e^{-{\beta} H}}/{{\rm tr}\left[ e^{-{\beta} H}\right]}.$ 
Furthermore exploiting the fact that 
\begin{equation}
{\rm det} [\mathbb{S}_{r, \phi}]={\rm det} [\mathbb{S}^T_{r, \phi}]=1~,
\end{equation}
one can extract {the state inverse temperature} ${{\beta}}$ of the state $\rho_G$  using the following relation
\begin{equation}
\nu_{{\beta}}=\sqrt{{\rm det} [\sigma]}~.
\end{equation}
Another quantity of interest -- see Fig.~(\ref{fig:ho}) (b) --- is the mean excitation number of a Gaussian state, whose expression in terms of the parameters $({\beta},~ {\pmb{\xi}},~ \chi)$ reads as \cite{lorch2018optimalA}
\begin{eqnarray}
\label{eq:mean-excitation-number-gaussian}
\langle a^\dag a \rangle=\frac{1}{2}\{\cosh(2 |\chi|) [2 N_b({\beta})+1] + {|\pmb{\xi}|}^2 - 1\}~.
\end{eqnarray}

\subsubsection{Dynamical Evolution} 
\begin{figure}
\vspace{.3cm}
\begin{overpic}[width=0.7\linewidth]{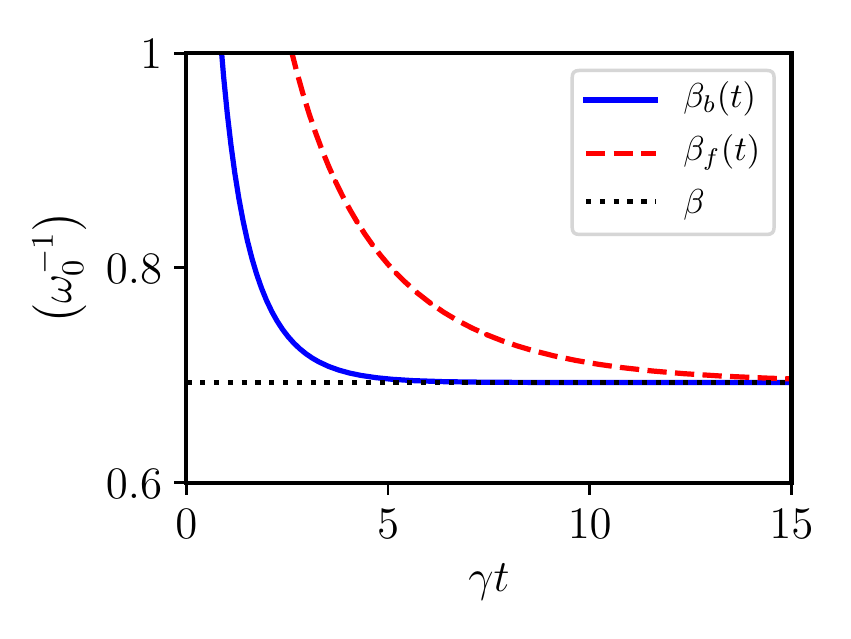}\put(12,95){} \end{overpic}
\caption{
\label{fig:betas}
(Color online)
Inverse temperatures (in units $1/\omega_0$) of the two Gaussian states as function of time: 
$\beta_b(t)$ (blue full line),
$\beta_f(t)$ (red dashed line),
choosing  the ground-state (i.e. $\beta_0\rightarrow \infty$) as initial state and $N_b(\beta)=1~.$
%
} 
\end{figure}

The ME we are considering in Eq.~(\ref{eq:lindblad-eq}) induces a Gaussian  mapping, meaning that 
it transform Gaussian states into other Gaussian states: namely
the time evolution from time $0$ to time $t~,$ dictated by Eq.~(\ref{eq:lindblad-eq}), simply maps 
$$\rho_G(\beta_0, {\pmb{\xi}}_0, \chi_0)\rightarrow \rho_G(\beta_q(t), {\pmb{\xi}}_q(t), \chi_q(t)),$$ where $q \in \{b, f\}$ is again the bath label.
To retrieve the explicit  temporal dependence of the quantities
$\beta_q(t)$, ${\pmb{\xi}}_q(t)$, $\chi_q(t)$ from the dynamical expression for the first and second moments of the ladder operators one can follow the 
same path we have detailed in the previous section to link $\beta, {\pmb{\xi}}, \chi$ to 
$\pmb{R}$ and $\sigma$.  
Finally, the same machinery can be used to relate the initial conditions to the parameters
of the input state giving 
\begin{eqnarray}
\langle a(0) \rangle &=&  A_1 (0)~, 
\\
\langle a^2(0) \rangle &=& \frac{1}{2}{\sigma_A}_{11}(0) + {A_1 (0)}^2~,
\\
\langle a^\dag a (0)\rangle &=& \frac{1}{2}[{\sigma_A}_{12}(0)-1] + {|A_1 (0)|}^2~,  
\end{eqnarray}
with
\begin{eqnarray}
\pmb{A}(0)&=&U^\dag \pmb{\xi}_0 ~,\\
\sigma_A(0)&=&\nu_{\beta_0} U^\dag  \mathbb{S}_{\chi_0} \mathbb{S}^T_{\chi_0} U^*
~.
\end{eqnarray}
and eventually one can monitor the initial mean excitation number by applying Eq.~(\ref{eq:mean-excitation-number-gaussian}) to the initial state.
As an application of this approach in Fig.~\ref{fig:betas} we report the values of $\beta_b(t)$ and $\beta_f(t)$ obtained by solving  Eq.~(\ref{eq:lindblad-eq}): 
in both cases we notice that dynamics  send asymptotically the system temperature $1/\beta_q(t)$ --- initially being $1/\beta_0$ --- to the bath temperature $1/\beta$, but with different rates, the slowest being the fermionic one.

\subsubsection{The quantum Chernoff quantity}

Let now $\rho_q(t)$ the Gaussian state (\ref{eq:gaussian-state}) of parameters  
${\beta}_q(t), {\pmb{\xi}}_q(t), \chi_q(t)$ describing the evolution of the density matrix (\ref{INPUTGAUS})
of the main text under the action of the ME (\ref{eq:lindblad-eq}) associated with the $q\in \{ b, f\}$ environment 
scenario. 
Following Ref.~\cite{calsamiglia2008quantumA} we can compute the value of the 
Chernoff quantity $Q_r(t)$ (\ref{eq:chernoff-bound})  via the expression 
\begin{equation}
\label{CB-formula}
Q_r(t)=
\frac{2 ~\mathcal{N}_{\beta_b, r}~ \mathcal{N}_{\beta_f, 1-r} ~
e^{ -{\pmb{\delta}}^T\left[\tilde{\sigma}_b(r)+\tilde{\sigma}_f(1-r)\right]^{-1}\pmb{\delta} }}{
\sqrt{\det\left[\tilde{\sigma}_b(r)+\tilde{\sigma}_f(1-r)\right]}}
~,
\end{equation}
where
$\pmb{\delta}=\pmb{\xi}_b-\pmb{\xi}_f $ 
is the difference between the first moments of the two states;
${\nu_{\beta_q}=\coth({\beta_q} \omega_0/2)=\sqrt{{\rm det}[ \sigma_q]}~;}$ 
$\mathcal{N}_{\beta_q, r}=\frac{(1-e^{-\beta_q \omega_0})^r}{1-e^{-\beta_q \omega_0 r}}$;
$\tilde{\sigma}_q(r)=\frac{\nu_{r  \beta_q}}{\nu_{\beta_q}} \sigma_q ~$
and $\sigma_q$ is the covariance matrix \cite{serafini2017quantumA}.

%
%
When the initial state has zero squeezing $(\chi_0=0)$
Eq.~(\ref{CB-formula}) assumes the simplified form 
\begin{eqnarray}
\label{eq:chernoff-displ-thermal-1}
Q(r, t)=
\frac{2~ \mathcal{N}_{\beta_b, r}~ \mathcal{N}_{\beta_f, 1-r} ~
}{
\nu_{\beta_b r}+\nu_{\beta_f (1-r)}}
~
e^{-\frac{|\pmb{\delta}|^2}{\nu_{\beta_b r}+\nu_{\beta_f (1-r)}}}
~,
\end{eqnarray}
with 
\begin{eqnarray}
\label{eq:chernoff-displ-thermal-coefficients}
\begin{cases}  
|\pmb{\delta}|^2=
|\pmb{\xi}_0|^2 
\left(e^{-\frac{\gamma}{2} t}-e^{-{\frac{\gamma}{2} t}/{ n_{\rm th}}}\right)^2
~,
\\
\nu_{\beta_q r}=2 \frac{1}
{\left[
{1}/{{N_b({\beta_{q}})}}+1 
\right]^r-1}+1
~,
\\
\mathcal{N}_{\beta_q, r}=
\frac{1}
{\left[1+{N_b({\beta_{q}})}\right]^r-\left[{N_b({\beta_{q}})}\right]^r}
~,
\\
{N_b({\beta_{q}})}=N_b(\beta_0)  e^{-\gamma / n_{\rm th}^{\rm (q)} t }  + N_b(\beta) \left(1-e^{-\gamma / n_{\rm th}^{\rm (q)} t }\right)
~.
\end{cases}
\end{eqnarray}
Notice that if we take initial state of the probe to be the ground state, i.e. 
${\pmb{\xi}}_0=0$ and $\beta_0\rightarrow \infty$,
the first moments vanish, i.e.  $\pmb{\delta}(t)=0$, and 
${N_b({\beta_{q}})}=N_b(\beta) \left(1- e^{-\gamma / n_{\rm th}^{\rm (q)} t }\right)~.$
For $\beta=\beta_0$ instead the above expression reduces to the one 
reported in the main text. 

\section{Details on TLS states}

The equation of motion for the TLS is obtained by considering
$\zeta_{p=TLS} = \sigma_{-}$ in Eq.~(\ref{eq:detailed-balance-2}), obtaining
\begin{equation}  \frac{d}{dt} \langle \sigma_+ \sigma_-(t) \rangle_q = -\gamma \frac{N_q(\beta)}{N_f(\beta)} \langle \sigma_+ \sigma_-(t) \rangle_q + \gamma N_f(\beta),  \end{equation}
from which  we immediately get
\begin{equation} \label{eq:pop} \frac{d}{dt} \langle \sigma_z (t) \rangle_q = -\gamma n_{\rm th}^{(q)} (\langle \sigma_z (t) \rangle_q - 2N_f(\beta) + 1),      \end{equation}
with $n_{\rm th}^{(q)}$ as defined in the main text. 
Accordingly the population of the TLS is always expected to equilibrate to $N_f(\beta)$, while the thermalization rates depends on the nature of the external bath.
On the contrary for the coherence terms we 
from (\ref{eq:lindblad-eq-general}) we get 
\begin{equation} \frac{d}{dt} \langle\sigma_x(t) \rangle_q = -\frac{\gamma n_{\rm th}^{(q)}}{2} \langle\sigma_x(t) \rangle_q ,\label{eq:coherence} \end{equation}
which can be easily integrated. 
The Helstrom error probability (\ref{Helstrom}) depends on the trace distance between
$\rho_{b}(t)$ and $\rho_{f}(t)$ that for a two level system reads
\begin{eqnarray}&& \label{eq:trdistq}  ||\rho_{b}(t) - \rho_{f} (t)||_1\nonumber \\
&&=  \sqrt{(\langle \sigma _x(t)\rangle_b -\langle \sigma_x(t) \rangle_f)^2 + (\langle \sigma _z(t)\rangle_b -\langle \sigma_z(t) \rangle_f)^2 }, \end{eqnarray}
where we supposed, without loss of generality, the $y$-component of the Bloch vector to be $0$ during all the process.
In comparison, the less straightforward equation (\ref{eq:chernq}) holds for the Chernoff quantity in the TLS case.

\subsubsection{Minimization of the trace norm in a TLS}

From the analysis of the previous section the square of the trace norm
for the bath tagging problem can be written as
\begin{eqnarray} \label{para1} &&\|\rho_b(t) - \rho_f(t)\|_1^2 \\
 &&\quad = (1-\langle \sigma_z(0) \rangle^2) f(t) + (\langle \sigma_z(0) \rangle - \langle \sigma_z \rangle_{eq})^2 g(t)\;,  \nonumber \end{eqnarray}
where we defined
$f(t) := (e^{\frac{-\gamma n^{(f)}_{th} t}{2}} - e^{-\frac{ \gamma  n^{(b)}_{th} t}{2}})^2$, $g(t) :=  (e^{-\gamma n^{(f)}_{th} t} - e^{-\gamma  n^{(b)}_{th} t})^2$, $ \langle \sigma_z \rangle_{eq} := 2N_f(\beta) -1 $
and used that, for an initial pure preparation, $\langle \sigma_x(0) \rangle^2 = 1- \langle \sigma_z(0) \rangle^2$.
The expression (\ref{para1}) is parabolic in $\langle \sigma_z(0) \rangle$ and
can be written in standard form as
\begin{equation} \begin{gathered} \label{Ydef} Y(\langle \sigma_z(0) \rangle, t) = \langle \sigma_z(0) \rangle^2(g(t)-f(t)) \\ - 2 \langle \sigma_z(0) \rangle \langle \sigma_z \rangle_{eq} g(t)+f(t) +\langle \sigma_z \rangle_{eq}^2 g(t).  \end{gathered} \end{equation}
Since we are interested in the maxima of $Y(\langle \sigma_z(0) \rangle, t)$ in the interval $-1 \leq \langle \sigma_z(0) \rangle \leq 1$ there are three possible candidates, i.e. the
vertex of the parabola and the two values at the extrema $Y(-1, t)$ and $Y(1,t)$.

The following two conditions are necessary for the vertex to be an acceptable maximum
\begin{enumerate}
 \item The concavity of the parabola has to be negative, that happens, from the (\ref{Ydef}), when $g(t)-f(t)\leq 0$;
 \item The abscissa of the vertex corresponds to a physical state, {\it i.e.} lies in the $[-1,1]$ interval. 
 More explicitly we have $-1 \leq \frac{ g(t) \langle \sigma_z \rangle_{eq}}{g(t)-f(t)} \leq 1$.
\end{enumerate}
Notice that since $ g(t)  \langle \sigma_z \rangle_{eq}  \leq 0$ and the first condition provides 
$g(t)-f(t)\leq 0$ the constraint on the abscissa of the vertex can be simplyfied to $ g(t) \langle \sigma_z \rangle_{eq}  \geq g(t)-f(t)$ that 
provides a stricter condition in respect to $g(t)-f(t) \leq 0$.
Explicitly solving the inequality  $g(t)  \langle \sigma_z \rangle_{eq}  \geq g(t)-f(t)$ we find that it holds for $t \geq t^*$, with $t^*$ such that
\ba e^{\frac{-\gamma n^{(f)}_{th} t^*}{2}} + e^{-\frac{ \gamma  n^{(b)}_{th} t^*}{2}} = {1}/{\sqrt{2 - 2 N_f(\beta)}}.\ea
It remains to compare $Y(-1, t)$ and $Y(1,t)$ when $t < t^*$ ({\it i.e.} the region in which the maximum is located
at the boundaries), with the ordinate of the vertex
$V(t) = f(t) - \langle \sigma_z \rangle _{eq}^2 \frac{f(t) g(t)}{g(t)-f(t)}$ computed in the part of the domain for which $t\geq t^*$.
For this sake we notice that for $t=t^*$ the ordinate of the vertex is exactly equal (by definition) to $Y(1,t^*)$ and 
that $V(t)$ is a decreasing function of $t$ in the region of interest {\it i.e.} $ V(t) \leq V(t^*)$ $\forall t \geq t^*$.
With this last argument we conclude that for all values of $t $ the function $V(t)$ is upper bounded by
$Y(1, t^*)$, proving in that way that the vertex is not the absoulute maximum,
that therefore lies wheter in $\langle \sigma_z(0) \rangle =1$ or $\langle \sigma_z(0) \rangle=-1$.
Is easy to show, again studying the properties of the parabolic function (\ref{para1}), that the value in $\langle \sigma_z(0) \rangle=1$ is always greater than
its opposite $\langle \sigma_z(0) \rangle=-1$, indeed
$Y(1,t) - Y(-1,t) = - 4 g(t) \langle \sigma_z \rangle_{eq} \geq 0$.
Thus we can plug $\langle \sigma_z(0) \rangle =1$ in the Eq. (\ref{para1}) obtaining
\begin{equation} Y(1, t) = (1-\langle \sigma_z \rangle_{eq})^2 g(t) , \end{equation}
that is exactly the square of the right hand side of the expression reported in the main text
and can be now studied as a function of the single parameter $t$.
Deriving this last equation and finding the root we obtain eq. (\ref{eq:bartqub})
that represent a local maximum in $t$, since $Y(1,t)$ is positive and nullifies at the extrema of the time domain 
(before starting the process and after a complete thermalization the two hypotheses are indistinguishable).

\end{document}